\documentstyle[11pt,a4]{article}

\renewcommand{\thefootnote}{\fnsymbol{footnote}}
\newcommand{\beq}{\begin{equation}}
\newcommand{\eeq}{\end{equation}}
\begin{document}
\thispagestyle{empty}
\begin{flushright}
UUITP-5/95\\
hep-th/9505009\\
\end{flushright}

\vspace{1cm}
\begin{center}

{\Large{\bf Comments on the Covariant $Sp(2)$-Symmetric
\\Lagrangian BRST Formalism} }
\end{center}
\vspace{2cm}
\begin{center}
A. Nersessian{\footnote{On leave of absence from the
Laboratory of Theoretical Physics, Joint Institute for Nuclear
Research, Dubna, 141980 Russia}}{\footnote{e-mail:
nerses@thsun1.jinr.dubna.su}} and
{}~P.H. Damgaard{\footnote{On leave of absence from the Niels Bohr
Institute, Blegdamsvej 17, DK-2100 Copenhagen, Denmark.}}\\
\vspace{1cm}
{\it Institute of Theoretical Physics, Uppsala University, Sweden}
\end{center}
\begin{abstract}
We give a simple geometrical picture of the
basic structures of the covariant $Sp(2)$ symmetric quantization
formalism -- triplectic quantization -- recently suggested by
Batalin, Marnelius and Semikhatov. In particular,
we show that the appearance of an even Poisson bracket is not a
particular property of triplectic quantization. Rather, any solution of
the classical master equation generates on a Lagrangian surface of the
antibracket an even Poisson bracket. Also other features of triplectic
quantization can be identified with aspects of conventional Lagrangian
BRST quantization without extended BRST symmetry.

\end{abstract}

\vfill
\setcounter{page}0
\renewcommand{\thefootnote}{\arabic{footnote}}
\setcounter{footnote}0
\thispagestyle{empty}

\newpage
\setcounter{equation}0
\section{Introduction}

An $Sp(2)$-symmetric Lagrangian BRST quantization presecription
reminiscent
of the conventional Batalin-Vilkovisky formalism \cite{BV} has been
known for a few years \cite{BLT}. Just as ordinary Batalin-Vilkovisky
Lagrangian quantization can be derived from the underlying principle
of imposing the Schwinger-Dyson BRST symmetry \cite{AD1} at all stages
of the quantization procedure (and thus ensuring correct Schwinger-Dyson
equations at the level of BRST Ward Identities) \cite{AD2}, the
$Sp(2)$-symmetric scheme of ref. \cite{BLT} can be derived from imposing
the $Sp(2)$-symmetric version of the Schwinger-Dyson BRST symmetry
\cite{DDJ}. See also the alternative derivation in \cite{DJ}.

Very recently, Batalin and Marnelius \cite{BM} have proposed a modified
$Sp(2)$-symmetric scheme, which reproduces all results of the earlier
method of ref. \cite{BLT}. The main advantage of the new proposal
is that it can be readily generalized to a covariant formulation,
a task carried out by Batalin, Marnelius and Semikhatov \cite{BMS}.
Here, ``covariant" refers to the supermanifold of fields (including all
necessary ghosts, auxiliary fields, etc.) and antifields (and even
further fields, see below). They have called this new formulation
``triplectic quantization". On top of the usual doubling of fields
(by the introduction of antifields), triplectic quantization involves
an additional tripling. For $N$ fields $\phi^A$, the dimension of the
supermanifold in question is thus $6N$.

In contrast to covariant formulations of conventional Batalin-Vilkovisky
quantization \cite{cov}, the covariant $Sp(2)$-symmetric formalism of
ref. \cite{BMS} involves a number of new complications that makes it
quite involved and which to some extent obscure its geometric meaning.
It may also appear surprising that in its precise formulation, the
triplectic formalism does not include the minimal solution found
in ref. \cite{BLT}.

The purpose of the present short letter is to provide
some observations which we believe will make this new covariant
$Sp(2)$-symmetric formalism more transparent. In so doing, we shall
demonstrate that several of the new features of the covariant
$Sp(2)$-symmetric quantization scheme have direct analogues also in
conventional Lagrangian BRST quantization (without this extended
BRST symmetry). We shall also show how the conditions for
$Sp(2)$-symmetric quantization can be generalized in a simple manner
to include the formulation of Batalin, Lavrov and Tyutin \cite{BLT}.

The main ingredients in triplectic quantization are a pair of
antibrackets $(\cdot,\cdot)^a$, a pair of operators $\Delta^a$, and
a pair of odd vector fields $V^a$. A number of consistency
conditions involving these objects need to be satisfied. These will
be reviewed below. It turns out that these consistency conditions
can be compactly formulated in terms of a conventional antibracket
defined on the $Sp(2)$-enlarged set of fields and antifields.

\section{A Poisson bracket from the antibracket}

We start with the conventional antibracket formalism, without at first
imposing the additional requirement of $Sp(2)$ symmetry. Also, we will
discuss the most general setting, not necessarily restricted to
Darboux coordinates. Thus, let ${\cal M}_0$ be a $2N$-dimensional
supermanifold with local coordinates $x^A$.
The bilinear differential operation
\beq
(f, g) = \frac{\partial^r f}{\partial x^A}(x^A,x^B)\frac{\partial^l f}
{\partial x^B}
\eeq
from differentiable functions $f(x)$ and $g(x)$ on ${\cal M}_0$ defines
the antibracket if it in addition satisfies the following 3 conditions.
First, it changes Grassmann parity:
\beq
\epsilon((f,g)) = \epsilon(f) + \epsilon(g) + 1 ~.
\eeq
Second, it satisfies the exchange relation
\beq
(f,g) = -(-1)^{(\epsilon(f)+1)(\epsilon(g)+1)}(g,f) ~,
\eeq
and, third, it fulfills the generalized Jacobi identity
\beq
(-1)^{(\epsilon(f)+1)(\epsilon(h)+1)}(f,(g,h)) + ~
{\mbox{\rm cycl. perm.}}~ = 0 ~.
\eeq

If in addition one has a volume element $[dv] = \rho(x)[dx]$
defined on ${\cal M}$, then one can introduce a
generalized $\Delta$-operator in covariant form \cite{cov}:
\begin{equation}
\Delta f =\frac{1}{2}{\mbox{\rm div}}_{\rho}(f,\cdot)~
\equiv~ \frac{1}{2}\frac{{\cal L}_{(f,\cdot)} dv}{dv}, \label{delta}
\end{equation}
where   ${\cal L}_{(f,\cdot)}$
denotes the Lie derivative along the
anti-Hamiltonian vector field $ (f,\cdot)$.
Locally, the $\Delta$-operator (\ref{delta}) takes the form
\begin{equation}
           \Delta f=\frac{1}{2}
     \frac{\partial^r}{\partial x^A}
(x^A,f) + \frac{1}{2\rho}(\rho,f) ~.
   \label{deltaloc}
\end{equation}

Some basic relations between the antibracket and the generalized
$\Delta$-operator follows straightforwardly from the Leibnitz rule
and the Jacobi identity:
\begin{eqnarray}
(-1)^{\epsilon(g)}(f,g) &=& \Delta(fg) - f\Delta g
 -(-1)^{\epsilon(g)}(\Delta f)g  \label{liebdelta}  \\
\Delta(f,g) &=& (f,\Delta g)
+(-1)^{\epsilon(g)+1}(\Delta f,g) ~.    \label{deltajac}
\end{eqnarray}

Consider next two volume elements $[d{\tilde v}]$ and $[dv]$
which are related by
$[d{\tilde v}] ={\rm e}^{2S(x)}[dv]$ (with $\epsilon(S)=0$). The
${\tilde\Delta}$-operator associated with $[d\tilde{v}]$ is
related to the $\Delta$-operator as follows:
\begin{eqnarray}
{\tilde\Delta} f& =& \Delta f + (S,f) =
{\rm e}^{-S}\Delta ({\rm e}^{S} f), \label{con} \\
{\tilde\Delta}^2 f &=& \Delta^{2} f  +
({\rm e}^{-S}\Delta {\rm e}^S, f) ~.
 \label{deltasqcon}
\end{eqnarray}
{}From these relations follow, particulary, that the squares af the
$\Delta$-operators coincide if the function $S(x)$ satisfies the
``quantum Master Equation" :
\begin{equation}
\Delta {\rm e}^{S} =0 ~~\Leftrightarrow~~ \Delta S +\frac{1}{2}
(S,S) = 0.\label{q}
\end{equation}
So the generalized $\Delta$-operator is not nilpotent with an
arbitrary volume element. We will assume that the nilpotency condition
holds with respect to the volume element $dv$. Assuming in addition
that $\Delta$ is a 2nd order operator, one can define the antibracket
according to eq. (\ref{liebdelta}).

After these general remarks, we are now ready to show how
one can define an ordinary even Poisson bracket from the antibracket.
To this end, assume that in addition to the above objects we can
supply an odd, nilpotent, vector field $V$ which
anticommutes with $\Delta$:
\beq
V\Delta = - \Delta V ~.
\label{anticom}
\eeq
It follows that this vector field differentiates the antibracket as:
\beq
V(f,g) = (Vf,g) + (-1)^{\epsilon(f)+1}(f,Vg) ~.
\label{diffV}
\eeq

Furthermore, let $u(x), v(x), w(x),...$ be functions on ${\cal M}$,
which commute with respect to the antibracket:
\begin{equation}
(u,v) = 0.
\label{g}
\end{equation}
Consider on this set of functions the operation
\begin{equation}
 \{u,w\} \equiv (u, Vw) ~.
\label{even}
\end{equation}
It is easy to check, using  $V^2=0$, eq. (\ref{diffV}) and
the Jacobi identity, that this operation satisfies
\begin{eqnarray}
(-1)^{\epsilon(u)\epsilon(v)}\{u,v\} & = & \{v,u\}  \nonumber\\
(-1)^{\epsilon(u)\epsilon(w)}\{\{u,v\},w\}  +
{\rm cycl.}\;{\rm perm.} & =& 0 ~.
\end{eqnarray}
Furthermore, using the derivation property of the antibracket it is
straightforward to see that the bracket (\ref{even}) satisfies the
Leibnitz rule
\beq
\{u,vw\} = \{u,v\}w + (-1)^{\epsilon(u)\epsilon(v)}v\{u,w\} ~.
\eeq

The expression (\ref{even}) thus defines an {\it even Poisson
bracket}-like operation on the set functions that satisfy (\ref{g}).
Particularly important realizations of the $V$-fields are Hamiltonian
vector fields (with respect to the antibracket), generated by solutions
of the classical Master Equation:
\beq
V = (S_0, \cdot) ~; ~~~~~(S_0,S_0) = 0~, ~~~~~\epsilon(S_0) = 0 ~.
\label{VHam}
\eeq
For any non-degenerate antibracket an arbitrary $V$-field can
at least locally be put in such a Hamiltonian form.
The classical Master Equation thus generates an even Poisson
bracket on an arbitrary isotropic surface of a non-degenerate antibracket.

\setcounter{equation}{0}
\section{The triplectic formalism}

We now turn to the main ingredients of the triplectic quantization scheme
proposed in refs. \cite{BM,BMS}.
In this scheme one requires the existance of a pair of $\Delta$-operators
$\Delta^a$ and odd vector fields $V^a$ ($a = 1, 2$) satisfying the
following consistency conditions:
\begin{eqnarray}
&\Delta^{\{a}\Delta^{b\}} =0,& \quad \epsilon(\Delta^a)=1\label{dd} \\
&V^{\{a}V^{b\}} =0,&\quad \epsilon(V^a)=1 \label{vv}\\
&V^a\Delta^b +\Delta^bV^a =0 &.
\label{vd}
\end{eqnarray}
Here and in the following the curly bracket denote symmetrization with
respect to the indices $a$ and $b$.
The operators $\Delta^a$ generate, due to (\ref{liebdelta}), a pair of
antibrackets $(\cdot,\cdot)^a$. The above consistency conditions
then imply:
\begin{eqnarray}
& &(-1)^{(\epsilon(f)+1)(\epsilon(h)+1)}(f,(g,h)^{\{a})^{b\}} + ~
{\mbox{\rm cycl. perm.}}~ = 0 \quad \label{bj}\\
& &\Delta^{\{a}(f,g)^{b\}} = (f,\Delta^{\{a} g)^{b\}}
+(-1)^{\epsilon(g)+1}(\Delta^{\{a} f,g)^{b\}} ~.    \label{ad}\\
& &V^a(f,g)^b = (V^af,g)^b  +(-1)^{\epsilon (f)+1} (f, V^a g)^b\label{av}
\end{eqnarray}

The partition function in the triplectic formalism
is defined by the expression
\begin{equation}
Z=\int [dv][d\lambda]{\rm e}^{\frac{i}{\hbar}\left[W(x) +X(x,\lambda)\right]}
{}~,\label{z}
\end{equation}
where $W(x)$ is viewed as the quantum action of the theory, and $X(x,\lambda)$
is considered the gauge fixing term. This division into two pieces of the
gauge fixed action is obviously to a large extent arbitrary, but we shall
follow the conventions of refs. \cite{BM,BMS}. Some background for this split
into $W$ and $X$ can be found in the recent work of Batalin and Tyutin
\cite{cov}. The gauge-fixing function $X$ restricts
the partition function to the ``space of effective fields" needed to describe
the quantum dynamics, and
$\lambda$ are a some additional parametric field variables. They are best
thought of as auxiliary fields in the path integral. They become simple
Lagrangian multipliers of gauge constraints when $X$ depends on them
linearly.

To define the pair of antibrackets, Batalin, Marnelius and Semikhatov
\cite{BM,BMS} extend the dimension of the
supermanifold from $(n|m)$ (the submanifold of
fields $\phi^A$) to $(2n+4m|2m+4n)$.

The partition function (\ref{z}) is gauge independent if the following
``quantum Master Equations" hold (using here
for convenience the formulation in which div$ V^a = 0$):
\begin{equation}
(\Delta^a + V^a){\rm e}^W = 0~,\quad
(\Delta^a-V^a + \ldots)
{\rm e}^X = 0~.
\label{2m}
\end{equation}
In the last equation the dots indicates the extra terms that are required
due to the variation of the parametric fields variables $\lambda$. These
terms are not of fundamental importance for the formalism\footnote{In
particular, they can be removed from the above Master Equations
by imposing an additional Master Equation condition on the $\lambda$-fields.},
and we shall therefore not display them in detail here. The crucial
difference between the Master Equation for $W$ and the analogous one for
$X$ is the sign in front of the ``transport term" induced by the vector
fields $V^a$.

We shall now provide an
interpretation of the consistency conditions imposed between the pair of
antibrackets $(\cdot,\cdot)^a$, the pair of $\Delta^a$ operators, and the
pair of odd vector fields $V^a, ~ a = 0, 1$. We shall also give a simple
picture of the r\^{o}le played by the vector fields $V^a$ in ensuring the
$Sp(2)$-invariant BRST symmetry. Once these vector fields $V^a$ are
appropriately interpreted, we will see that the emergence of an
additional Poisson bracket structure within this $Sp(2)$ invariant scheme
follows naturally from the Poisson bracket defined through the antibracket,
as in the previous section.

The central ingedients on which many of our
subsequent considerations are based, are the following:
Consider a $\Delta$-operator and a $V$-field which both depend on a
pair of real parameters $k_a$:
\begin{eqnarray}
\Delta_k &=& \sum_{a=0,1} k_a\Delta^a \label{dk}\\
V_k &=& \sum_{a=0,1} k_aV^a ~,
\label{keq}
\end{eqnarray}
with $k_a =$ const. and $\epsilon(k_a) = 0$. The antibracket generated
by eq. (\ref{liebdelta}) then splits into the sum
\beq
(f,g)_k ~=~ \sum_{a=0,1} k_a(f,g)^a ~.
\label{bi}
\eeq

Remarkably,
requiring the nilpotency condition $\Delta_k^2 = 0$ to hold for all
$k_a$ is equivalent to the condition (\ref{dd}), and similarly
the relation (\ref{liebdelta}) between (\ref{bi}) and (\ref{dk})
coincides with the condition (\ref{ad}). In addition,
from the Jacobi identity for the antibracket
(\ref{bi}) for all $k_a$ follows
the consistency condition (\ref{bj}) for the pair of antibrackets
in the triplectic formalism.

Now, from the fact that $V_k$ anticommutes with $\Delta_k$
(eq. (\ref{anticom})), we find
\beq
\Delta^{\{a} V^{b\}} + V^{\{a}\Delta^{b\}} = 0
\label{vdw}.
\eeq
It similarly follows from eq. (\ref{diffV}) that

\beq
V^{\{a}(f,g)^{b\}} = (V^{\{a}f,g)^{b\}}  +(-1)^{\epsilon (f)+1} (f, V^{\{a}
g)^{b\}} \label{avw}.
\eeq

It is also clear that the $V_k$-field generates a Poisson bracket
as described in the previous section.
The generation of such an even bracket on isotropic surfaces
of the antibracket is therefore not a special property of
triplectic quantization.

One important observation should be mentioned at this point.
The relations (\ref{avw}) are (\ref{vdw}) are more general
than the analogous (\ref{av}) and (\ref{vd}) of the triplectic
formalism proposed in ref. \cite{BMS}.\footnote{In ref. \cite{BMS} these
identities are required to hold even before symmetrization in the
$Sp(2)$ indices.} However,
all other relations which follow from the above construction coincide
with the ones of triplectic quantization.
In fact, these plus the more general conditions (\ref{avw}) and
(\ref{vdw}) seem
to be sufficient for the construction of an $Sp(2)$ symmetric
quantization formalism. The original scheme of Batalin, Lavrov and Tyutin
\cite{BLT} is indeed based on these more general conditions.

The extended BRST symmetry of the $Sp(2)$-symmetric partition function
(\ref{z}) now acquires a very simple interpretation. First, the combined
invariance under both BRST and  anti-BRST transformations \cite{BMS}
takes in the notation above the form
\begin{equation}
\delta x=\sum_{a=0,1}k_a[(W-X,x)^a +2V^a] = (W-X,x)_k + 2V_k ~.
\label{symmetry}
\end{equation}
Explicitly, invariance leads to the following condition:
\beq
[\Delta_kW+V_kW+(W,W)_k] + [\Delta_kX-V_kX+(X,X)_k] + {\mbox{\rm
div}} V_k = 0 ~.
\eeq
Thus, if the Master Equations (\ref{2m}) hold, and div$ V_k=0$, then
eq. (\ref{symmetry}) is a symmetry.

Second, consider the case where $V_k$-field is Hamiltonian with respect
to the antibracket (\ref{bi}), $i.e.$, has the form (\ref{VHam}) for some
solution $S_0$ of the classical Master Equation whish is independent
of $k_a$. We can then define
\begin{equation}{\tilde{W}}
\equiv W-S_0 ~,\quad {\tilde{X}} \equiv X+S_0 ~.
\label{tilde}
\end{equation}
The partition function (\ref{z}) is trivially invariant under the
replacement $X \to {\tilde{X}}$, $W\to{\tilde{W}}$. This simply reflects
the unavoidable ambiguity in defining in this way what is meant by the
``quantum action'' ($W$) and the ``gauge fixing'' ($X$) before boundary
conditions have been imposed. But the BRST symmetry of the gauge-fixed
partition function depends on the {\em difference} $W - X$, and this
would seem to imply that the BRST transformations should change under
the above substitutions. We will resolve this apparent contradiction
below.

The extended BRST symmetry (\ref{symmetry}) now acquires a very simple
interpretation. Namely, it takes in the above notation the form
\beq
\delta x = ({\tilde{W}} - {\tilde{X}},x)_k ~.
\label{symnew}
\eeq
Next, taking into account the last equality in eq. (\ref{con}), we can
rewrite the Master Equations (\ref{2m}) in the form
\begin{equation}
   \Delta_k {\rm e}^{{\tilde{W}}} = 0 ~,~\quad \Delta_k{\rm e}^
{{\tilde{X}}}=0 ~.
\label{m3}
\end{equation}

So the triplectic formalism can be brought in a form
analogous to the conventional Batalin-Vilkovisky quantization (when
expressed with the help of $W$ and $X$).
Of course, the inverse of this
statement does not hold, since the supermanifold in question in general
needs to be larger than that of conventional Batalin-Vilkovisky
quantization. However, there can exist instances where a hidden $Sp(2)$
symmetry is present even in the conventional space fields and antifields
within conventional Batalin-Vilkovisky quantization.

The expressions (\ref{m3}) and (\ref{symmetry}) formally coincide
with the corresponding ones in the generalized BV-formalism \cite{BMS}.
In this sense the triplectic formalism can be viewed as the special case
of the antibracket formalism in which the antibracket depends on a
parameter of the circle $S^1$, and where an additional solution $S_0$
of the classical Master Equation, independent of these parameters,
exists.

\section{A Related Formulation without $Sp(2)$ Symmetry}

At this point we would like to make more direct contact with
the conventional
Batalin-Vilkovisky Lagrangian quantization (without extended BRST symmetry).
It is suggested in the concluding remarks of ref. \cite{BMS} that there
might exist a formulation without $Sp(2)$ symmetry that still involves
a nilpotent odd vector field $V$ which differentiates the antibracket.
At first sight there may not seem to much room for such an extension
of the conventional Batalin-Vilkovisky formalism. As formulated in
ref. \cite{BV}, there is indeed no obvious candidate for such a new
vector field $V$, simply by ghost number conservation. However, the most
general formulation of Lagrangian BRST quantization which has been
derived in ref. \cite{AD2} only reduces to the conventional
Batalin-Vilkovisky formalism in the special case where the integration
over the ghost partners $c^A$ of the antighosts (now viewed as
``antifields") $\phi^*_A$ can be integrated out of the path integral
to leave a $\delta$-function constraint on the antighosts $\phi^*_A$.
If one leaves the fields $c^A$ in the extended action in full
generality, the quantum Master Equation for the quantum action
$S$ is in fact not the one of ref. \cite{BV}, but rather
(in Darboux coordinates) \cite{AD2}:
\beq
\frac{1}{2}(S,S) = -\frac{\delta^r S}{\delta\phi^A}c^A + i\hbar
\Delta S ~.
\eeq

Only when assuming the simple ansatz ${\tilde S}[\phi,\phi^*,c] =
S^{BV}[\phi,\phi^*] - \phi^*_Ac^A$
does this complete Master Equation reduce to the conventional
Batalin-Vilkovisky Master Equation for $S^{BV}$. Clearly,
\beq
V ~= (-1)^{\epsilon_A+1}c^A\frac{\delta^l}{\delta\phi^A}
\label{ADV}
\eeq
satisfies $V^2 = 0$. It also differentiates the antibracket in the
sense of eq. (\ref{diffV}),
and it anticommutes with the $\Delta$-operator as in eq (\ref{anticom}).

In this notation, the full quantum Master Equation \cite{AD2}
takes the form
\beq
\frac{1}{2}(S, S) = VS + i\hbar\Delta S ~,
\eeq
in perfect analogy with the $Sp(2)$-symmetric Master Equation
proposed in ref. \cite{BMS}. In hindsight, it is not at all
surprising that such a formulation exists, nor is it surprising
that one needs to enlarge (slightly) the set of fields (by keeping
the ghosts $c^A$ instead of integrating them out in the path
integral) in order to find it. The ghosts $c^A$ of ref. \cite{AD2}
are precisely what become the $Sp(2)$-symmetric partners in
an $Sp(2)$-invariant formulation. It is interesting that even
the ``third set of fields" ($\bar{\phi}_A$ in the notation
of \cite{BLT}) have a completely natural place in the conventional
Lagrangian BRST quantization scheme (without extended BRST symmetry).
They are simple linear combinations of the collective fields that
are needed to derive the Schwinger-Dyson BRST symmetry \cite{DDJ}
through shifts $\phi^A \to \phi^A - \varphi^A$ (where the fields
$\varphi^A$ are linear combinations of the fields $
\bar{\phi}_A$ that are required in the $Sp(2)$-symmetric
formulation).
In conventional Lagrangian BRST quantization one normally
integrates these fields out of the path integral. But one could
easily keep them, in which case the formalism of Batalin,
Marnelius and Semikhatov \cite{BM,BMS} would look even less
different from the quantization scheme that does not impose
$Sp(2)$ symmetry \cite{AD2}.

The ghosts $c^A$ are ``spectator fields" in the antibracket, and
there is likewise no need in conventional Lagrangian BRST
quantization for an antibracket associated with the collective
fields. Such an additional antibracket appears only if one insists that
even the collective fields themselves shall obey their
correct Schwinger-Dyson equations at the level of BRST Ward
Identities.
Continuing iteratively in this way, one can clearly double the
number of fields as many times one wishes, each time introducing
a new antibracket for the new collective fields. The final
outcome is clearly unaltered by such an unnecessary complication.
In the $Sp(2)$-symmetric formulation it is however imperative
that at least the ``first-stage" collective fields are kept.
Otherwise the Schwinger-Dyson BRST--anti-BRST operator is not
nilpotent even when just restricted to the set of fields $\phi^A$
\cite{DDJ}.

The vector field of eq. (\ref{ADV}) is generated by $S_0
= \phi^*_Ac^A$ in the sense that $V = (S_0,~\cdot~)$. Indeed,
the disappearance of this vector field from the Master Equation
when substituting the ansatz $S[\phi,\phi^*,c] = S^{BV}[\phi,\phi^*]
- S_0[\phi^*,c]$ can be understood in the same simple manner
as discussed in the previous section for the $Sp(2)$-symmetric case.
Again, subtracting the generator of $V$ from the action in general
changes the boundary conditions. In the formulation of ref. \cite{AD2}
this change in boundary conditions is automatically
avoided precisely because it
involves an additional set of ghost fields (the $c$'s). Of course, in
the formulation based on $S^{BV}$ one should not integrate over the
$c^A$-fields in the partition function.

Remaining within the framework of Darboux coordinates, we note that
the natural condition of the vector field $V$ being divergence-free,
\beq
{\mbox{\rm div~}} V ~=~ 0 ~,
\eeq
is the simple statement that the functional measure of the
$\phi^A$-fields is invariant under arbitrary local shifts (a trivial
but nevertheless
implicit assumption in conventional Batalin-Vilkovisky quantization
when restricted to Darboux coordinates). When the functional
measure is not invariant under such shifts, one can still follow
the procedure of \cite{AD2} through (see the last reference in
\cite{cov}), and one then derives the covariant version of
Batalin-Vilkovisky quantization. This covariant description also
includes the additional vector field $V$ in its most general
formulation.

\vspace{0.5cm}

\noindent
{\sc Acknowledgements:}~
We would like to
thank F. De Jonghe and A. M. Semikhatov for stimulating discussions.
A.N. is indebted to A. Niemi for the warm
hospitality at the Institute of Theoretical Physics, Uppsala University,
where this work was carried out. The work of A.N. was
supported in part by grant No. M2I000 from the International
Science Foundation.

 \end{document}